# H- Ion Source Development for the FNAL 750keV Injector Upgrade

D.S. Bollinger

<sup>a</sup>Fermi National Accelerator Laboratory, Box 500, MS 307, Batavia, Illinois, 60510

**Abstract.** The new FNAL 750keV injector upgrade for the replacement of the 40 year old Fermi National Laboratory (FNAL) Cockcroft-Walton accelerators with a new ion source and 200MHz Radio Frequency Quadrupole (RFQ), Low Energy Beam Transport (LEBT) and Medium Energy Beam Transport (MEBT) [1], has been built and is now being tested prior to installation during the 2012 shutdown. The new H- ion source is a round aperture magnetron which was developed at Brookhaven National Lab (BNL) by Jim Alessi[2]. Operational experience from BNL has shown that this type of source is more reliable with a longer lifetime (on the order of 6 to 9 months) due to better power efficiency. With a similar duty factor to BNL, we expect to have a comparable lifetime between source changes. The new source design reliably produces 90mA of H- beam current at 15Hz rep-rate, 250µs pulse width, and a duty factor of 0.38%.. The measured emittances at the end of the LEBT are horizontally $\varepsilon_H = 0.21\pi$ mm$_*$ mrad and vertically $\varepsilon_V = 0.17\pi$ mm$_*$ mrad. With 35kV extraction the power efficiency is 60mA/kW. The source design, along with data from a test stand and the LEBT, will be presented in this paper.

## INTRODUCTION

The FNAL 750keV injector upgrade is slated for installation during the 2012 shutdown pending successful performance verification of all accelerator components including the new magnetron ion source. The injector currently being tested is shown in Figure 1. All of the accelerator components currently exist except for the MEBT. Several diagnostic lines have been built and installed to help characterize the injector. These include emittance probes, fast and slow Faraday cups, time of flight, wire scanner, and a spectrometer. Over the past several months the beam energy out of the RFQ, transmission efficiency of the RFQ, both input and output emittance of the RFQ, and the Einzel Lens chopper rise and fall times have been measured.

Along with testing the ion source on the LEBT, the source has also been installed on a test stand to verify source performance. The test stand shown in Figure 2 includes an Einzel lens focusing element, toroid, Faraday cup, and vertical and horizontal emittance probes.

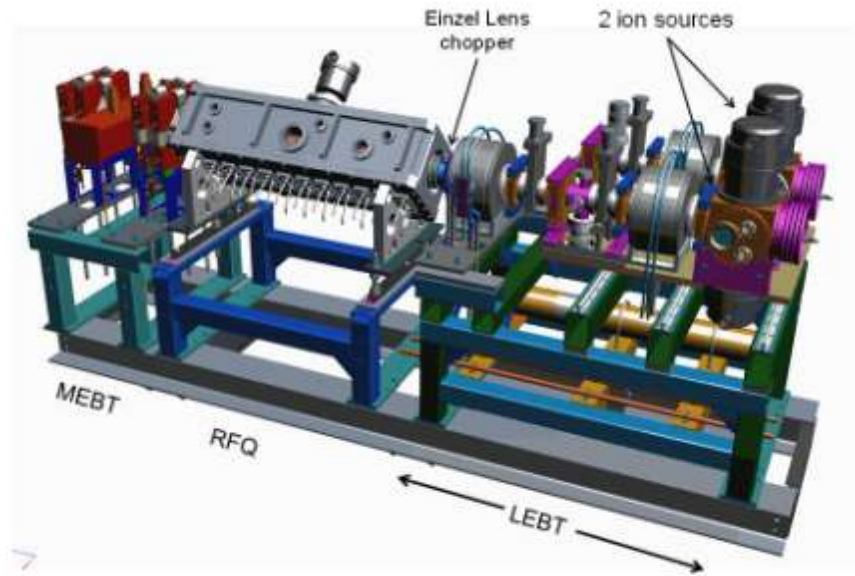

**Figure 1** New FNAL injector bemline under test

## SOURCE DESIGN

The source design is similar to both the BNL ion source and the source developed by CW Schmidt for the High Intensity Neutrino Source HINS project at FNAL[3]. The source shown in Figure 2, is mounted reentrant in a 10 inch vacuum "cube" and was designed with "ease of maintenance" in mind. It has a round aperture with a 45 degree extraction cone. The source cathode has a spherical dimple which provides focusing of the plasma to the anode aperture. The anode and extraction cone apertures are 3.175mm, which is what BNL currently uses. These apertures will be optimized to give the correct beam current and emittance for maximum efficiency of the Linac.

The extraction is single stage with the extraction cone at ground potential and the source pulsed to -35kV (the extractor is electrically connected to ground with finger stock, shown in the lower right picture of Figure 2). This extraction scheme allows the source to run in the space charge limited regime with high extracted beam currents. The extractor pulser is a new design that uses a vacuum tube as the switch. The source electronics reside in a floating high voltage rack which is pulsed to -35kV.

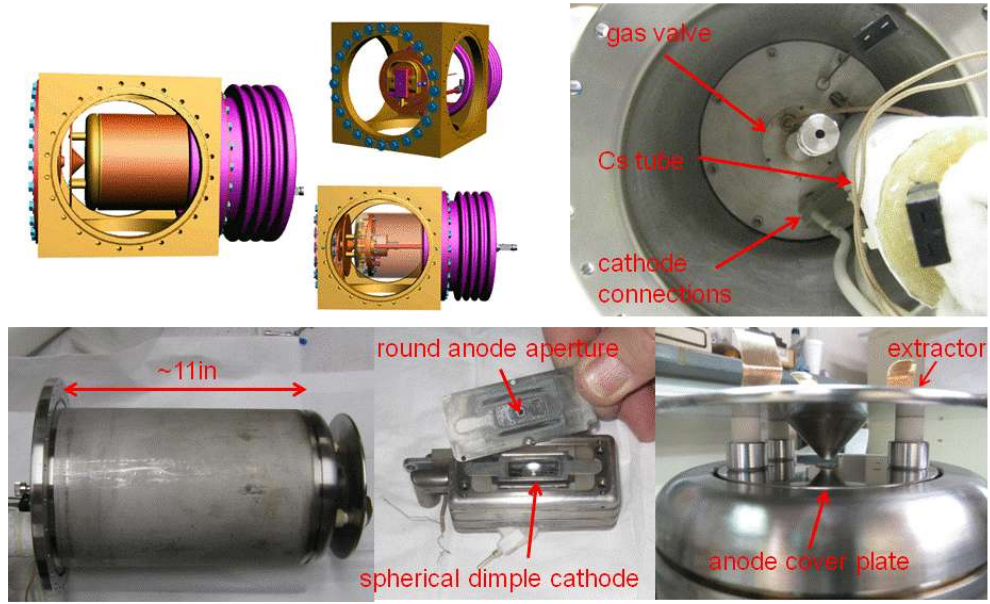

**Figure 2** New magnetron source with round aperture

## STUDIES

Initial source studies were performed using a modified HINS ion source. This source worked well for initial studies. However, the source had separate extraction and acceleration electrodes instead of a single stage extraction/acceleration like the BNL source. Problems arose when we tried to modify the source for single stage extraction/acceleration. The extractor had a high voltage feedthrough that was not needed. Attempts to eliminate this connector led to a lot of sparking related failures. At that time, it was decided to design and build a new source.

Several studies have been completed to characterize the new source. These include emittance measurements, beam perveance, electron to H⁻ ratio and, extraction gap spacing. These studies were designed to help us with future tuning and optimization of the source design.

## Beam Perveance

The beam current out of the source was measured in the test stand at several extraction voltages. Shown in figure 3 the current density at 80mA is $1A/cm^2$. The maximum beam current seen so far is 94mA at 35kV, which is comparable to the BNL source [2].

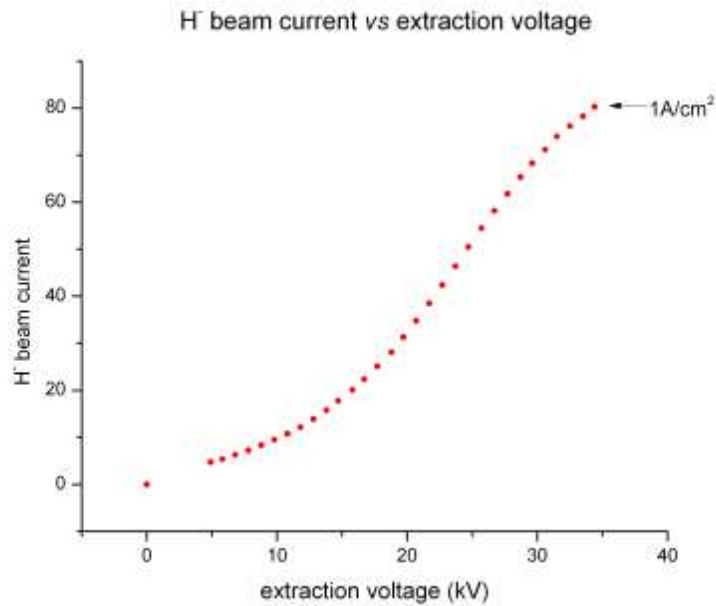

**Figure 3** Beam current vs extraction voltage

## Electron to H⁻ ratio

The electron to H⁻ ratio was measured on the test bench. The Faraday cup that normally resides on the far downstream end of the test stand was moved to the downstream aperture of the source cube. The Faraday cup aperture is 3 inches. SIMION simulations show that for an assumed 100mA, 35keV beam, the maximum distance from the source aperture that could be intercepted by the Faraday cup is 5.5 inches, which is within 1 inch of the source cube aperture.

A shunt resistor was installed in the extractor pulser that allowed the total extractor current to be measured. The extractor current was first measured with no source arc current. This was then used to normalize the current with beam.

The total current on the extractor is assumed to be the sum of the electrons and extracted H⁻ ions. Figure 4 shows a typical scope trace of the total extractor current.

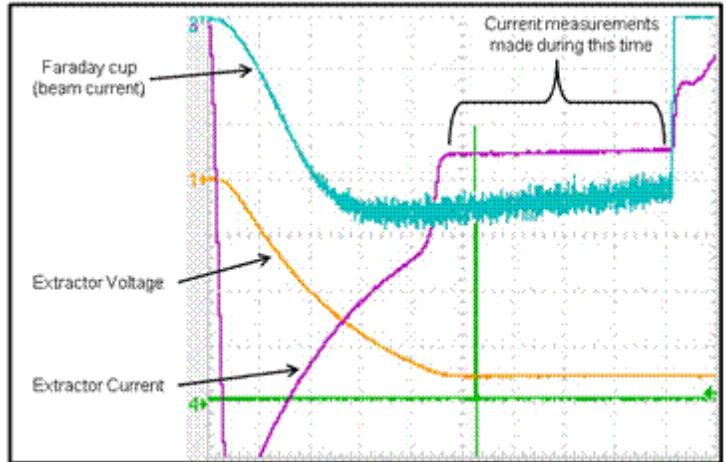
**Figure 4** total extractor current

The total current during the flattop, when there was H⁻ ions extracted for various extraction voltages is shown in figure 5. At 35kV the electron to H⁻ ratio is 1/1 with a source impedance of 11 Ohms.

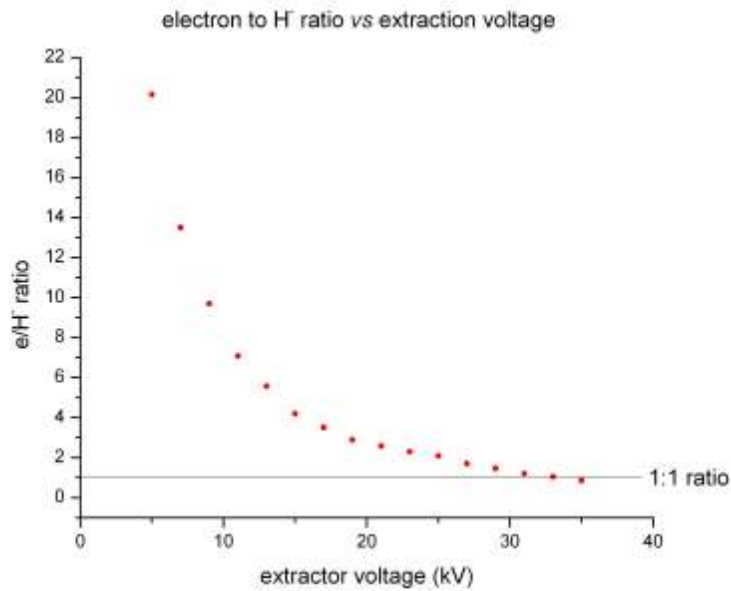
**Figure 5** Electron to H⁻ ratio for various extraction voltages

## Extraction Gap Studies

Various extraction gap sizes, ranging from 4.4mm to 4.9mm were studied to find a balance between spark rates and source performance. These studies were done on the test stand at reduced extraction voltages due to limitations in the stand optics. The first study was to see what the extracted beam current was for each of the gap sizes. As shown in Figure 6, the 4.4mm gap had the largest extracted beam current and the 4.9mm had the smallest.

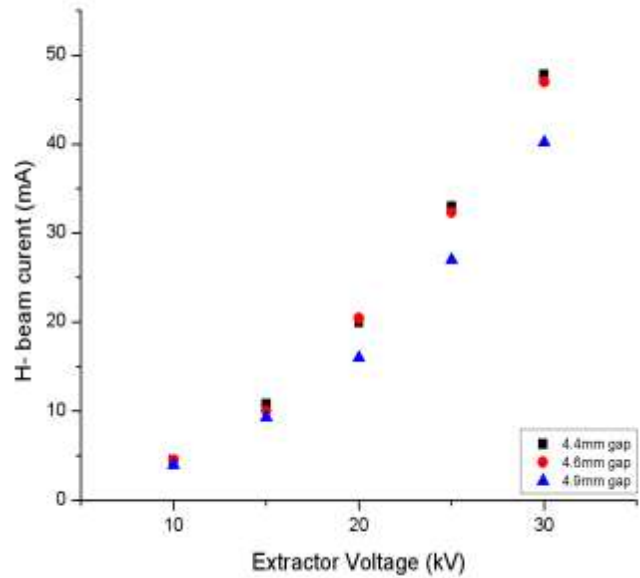

**Figure 6** Beam current vs extraction voltages for various extraction gap sizes

The second extraction gap study explored the beam emittance at several extraction voltages and beam currents. The data in Figure 7 suggest that the smallest extraction gap (4.4mm) has the largest emittance of the three sizes tried. The 4.6mm and 4.9mm gaps have roughly the same emittance (within the error of the measurement). This may be due to the lower extracted beam current.

While the source was installed in the LEBT for RFQ studies, it had an extraction gap of 4.6mm. To obtain the optimal beam current for RFQ studies the source needed to run at 11A of arc current. With the current pumping speed on the source cube (2000L/S for H2) the average pressure in the source cube was 2.7 X 10-6 Torr. With this pressure the spark rate was very high. The source was eventually removed, cleaned and the extraction gap was increased to 4.9mm. With this gap size there is almost no extractor sparking with source pressures up to 3.2 X 10-6 Torr. Based on this along with the emittance data, we plan to commission the RFQ after it is installed in the Linac with this extractor gap spacing.

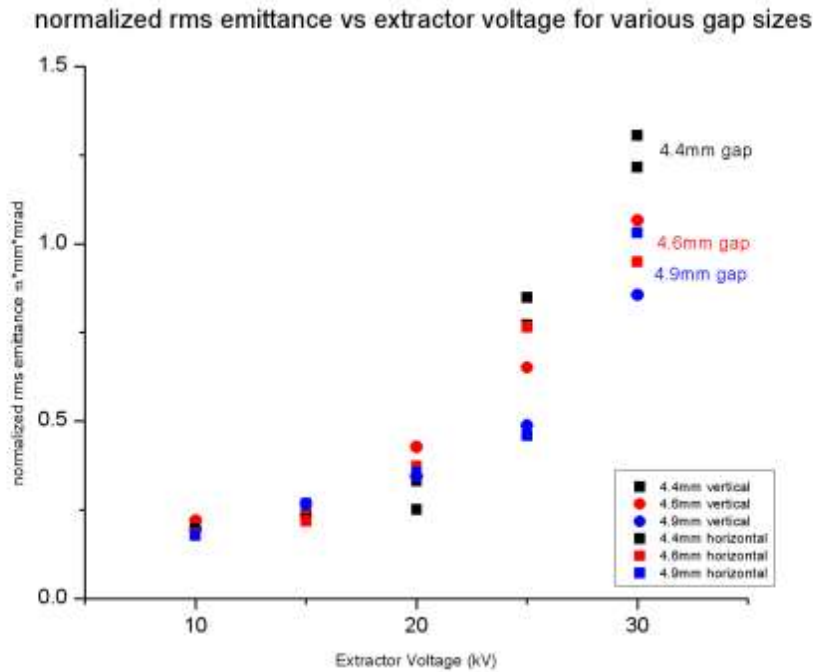

**Figure 7** Normalized emittance for various extraction gap sizes

## Emittance as a Function of Arc Current Study

The final study has involved emittance with respect to arc current. This study was motivated by a paper written by S.R. Lawrie, Dan Faircloth, A.P. Letchford, C.Gabor, and J.K. Pozmiski titled "Plasma meniscus and extraction electrode studies of the ISIS H- ion source"[4] The emittance was measured while the arc current was varied from 10A to 20A while the source was in the test stand and at the downstream end of the LEBT. The data from the test stand were taken with an extraction voltage < 30kV which is lower than required for injection into the RFQ. This data shown in Figure 8a, indicate that the lower the arc current the smaller the emittance. The difference between the horizontal and vertical planes is due to the source magnetic field in the extraction region. We do plan to explore either mounting the source at an angle or offsetting the cathode dimple to reduce this affect. The data taken while the source was mounted on the LEBT are shown in Figure 8b. These data are not consistent above 12A of arc current and are still being analyzed. These two sets of data do show that the average normalized rms emittance is smaller for lower arc currents. Also, the beam is more symmetrical at beam arc currents close to 10A. Coincidentally, RFQ transmission efficiency studies have shown that an arc current of 11A, yielded a transmission efficiency >70%. As a result we will use 11A of arc current for initial RFQ commissioning.

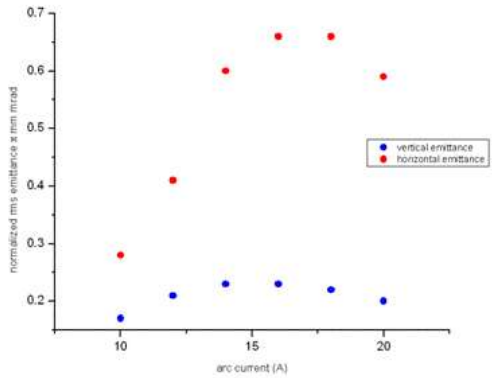 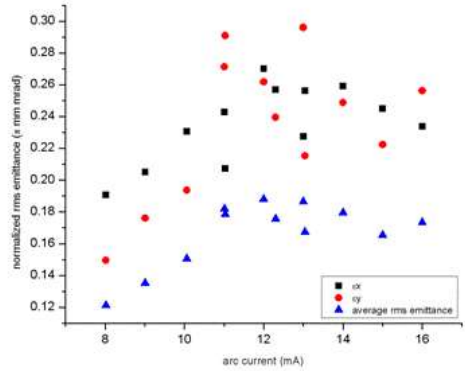

**Figure 8** Emittance as a function of arc current

# CONCLUSION

The current FNAL operational slit aperture ion sources are in the process of being replaced with round aperture magnetron sources similar to BNL. The new sources have been tested in both a test stand and on the new LEBT. The sources have been running for several months on the LEBT for both LEBT testing and RFQ testing. With this experience and the studies performed we have come up with the start up parameters listed in Table 1, that will be used for commissioning the new 750keV injector after it is connected to the Linac later this year.

**TABLE 1.** Start up parameters for new 750keV Injector commissioning

| Parameter | Value | Units |
|---|---|---|
| Arc Currrent | 11 | A |
| Arc Voltage | 150 | V |
| Extractor Voltage | 35 | kV |
| Beam Current | 80 | mA |
| Power Efficiency | 48 | mA/kW |
| Rep Rate | 15 | Hz |
| Arc Pulse Width | 250 | μs |
| Extracted Beam Pulse Width | 80 | μs |
| Duty Factor | 0.38 | % |
| Cathode Temperature | 380 | C |
| Cs Boiler Temperature | 130 | C |
| Emittance $\varepsilon x/\varepsilon y$ | 0.2 /0.3 | π mm mrad (normalized rms) |
| Extraction Gap | 4.9 | mm |

# ACKNOWLEDGMENTS

I would like to thank the FNAL Preacc Group and mechanical support technicians for their help with this project along with Pat Karns who helped take a lot of this data. I would also like to thank CY Tan for helping me with the data analysis.

# REFERENCES

1. C.Y. Tan, D.S. Bollinger, C.W. Schmidt, "750 keV LINAC INJECTOR UPGRADE PLAN" Beams-doc-3646-v, 2 2010.
2. J.G. Alessi, "Performance of the Magnetron $H^-$ Source on the BNL 200 MeV Linac", AIP Conf. Proc., Vol. 642, pf 279-286, 2002.
3. R. Webber, " $H^-$ Ion Source requirements for the HINS R&D Program", Beams-doc-3056-v1, 2008
4. S.R. Lawrie, Dan Faircloth, A.P. Letchford, C.Gabor, and J.K. Pozmiski "Plasma meniscus and extraction electrode studies of the ISIS $H^-$ ion source", Review of Scientific Instruments 81, 02A707, 2010